%% file: ifoDM.tex
\documentclass[prl, aps, reprint, nofootinbib, notitlepage, tightenlines,
preprintnumbers,showpacs,superscriptaddress]{revtex4-1}

\usepackage[hyperindex, breaklinks, colorlinks=true, citecolor=RoyalBlue, 
            linkcolor=BrickRed, urlcolor=ForestGreen]{hyperref}
\usepackage{graphicx,epstopdf}

\usepackage[usenames, dvipsnames, svgnames, table]{xcolor}

\usepackage{natbib,bm}
\usepackage{slashed}
\usepackage{hyperref}
\usepackage{breakurl}
\usepackage{multirow}
\usepackage{bbold}

\usepackage{mathtools, amssymb, amsfonts}
\usepackage{txfonts}

\def\be{\begin{equation}}
\def\ee{\end{equation}}
\def\ba{\begin{eqnarray}}
\def\ea{\end{eqnarray}}
\def\ge{\mathrel{\raise.3ex\hbox{$>$\kern-.75em\lower1ex\hbox{$\sim$}}}}
\def\la{\mathrel{\raise.3ex\hbox{$<$\kern-.75em\lower1ex\hbox{$\sim$}}}}

\def\simgt{\mathrel{\raise.3ex\hbox{$>$\kern-.75em\lower1ex\hbox{$\sim$}}}}
\def\simlt{\mathrel{\raise.3ex\hbox{$<$\kern-.75em\lower1ex\hbox{$\sim$}}}}

\newcommand{\nc}{\newcommand}

\nc{\gone}{\bar g_{\pi NN}^{(1)}}
\nc{\gzero}{\bar g_{\pi NN}^{(0)}}
\nc{\al}{\alpha}
\nc{\ga}{\gamma}
\nc{\de}{\delta}
\nc{\ep}{\epsilon}
\nc{\ze}{\zeta}
\nc{\et}{\eta}
\nc{\ka}{\kappa}
\nc{\rh}{\rho}
\nc{\si}{\sigma}
\nc{\ta}{\tau}
\nc{\up}{\upsilon}
\nc{\ph}{\phi}
\nc{\ch}{\chi}
\nc{\ps}{\psi}
\nc{\om}{\omega}
\nc{\Ga}{\Gamma}
\nc{\De}{\Delta}
\nc{\La}{\Lambda}
\nc{\Si}{\Sigma}
\nc{\Up}{\Upsilon}
\nc{\Ph}{\Phi}
\nc{\Ps}{\Psi}
\nc{\Om}{\Omega}
\nc{\ptl}{\partial}
\nc{\del}{\nabla}
\nc{\ov}{\overline}
\nc{\newcaption}[1]{\centerline{\parbox{15cm}{\caption{#1}}}}
\nc{\us}{U(1)$_S$}

\def\beq{\begin{equation}}
\def\eeq{\end{equation}}
\def\bmat{\begin{displaymath}}
\def\emat{\end{displaymath}}
\def\bear{\begin{eqnarray}}
\def\eear{\end{eqnarray}}
\def\ba{\begin{eqnarray}}
\def\ea{\end{eqnarray}}
\def\bery{\begin{array}}
\def\ery{\end{array}}
\def\bit{\begin{itemize}}
\def\eit{\end{itemize}}
\def\ben{\begin{enumerate}}
\def\een{\end{enumerate}}
\def\btab{\begin{tabular}}
\def\etab{\end{tabular}}
\def\btbl{\begin{table}}
\def\etbl{\end{table}}
\def\bfig{\begin{figure}[htb]}
\def\efig{\end{figure}}
\def\bpic{\begin{picture}}
\def\epic{\end{picture}}

\def\ga{\mathrel{\raise.3ex\hbox{$>$\kern-.75em\lower1ex\hbox{$\sim$}}}}
\def\la{\mathrel{\raise.3ex\hbox{$<$\kern-.75em\lower1ex\hbox{$\sim$}}}}
\def\gappeq{\mathrel{\rlap {\raise.5ex\hbox{$>$}}
{\lower.5ex\hbox{$\sim$}}}}
\def\lappeq{\mathrel{\rlap{\raise.5ex\hbox{$<$}}
{\lower.5ex\hbox{$\sim$}}}}

\def\gyr{{\rm \, G\kern-0.125em yr}}
\def\mev{{\rm \, Me\kern-0.125em V}}
\def\gev{{\rm \, Ge\kern-0.125em V}}
\def\tev{{\rm \, Te\kern-0.125em V}}

\begin{document}

\title{Laser Interferometers as Dark Matter Detectors}

\author{Evan D. Hall}
\affiliation{California Institute of Technology, Pasadena, CA 91125, USA}

\author{Thomas Callister}
\affiliation{California Institute of Technology, Pasadena, CA 91125, USA}

\author{Valery V. Frolov}
\affiliation{LIGO Livingston Observatory, Livingston, LA 70754, USA}

\author{Holger M\"uller }
\affiliation{Department of Physics, 366 Le Conte Hall, University of California, Berkeley, CA 94720,
USA}

\author{Maxim Pospelov}
\affiliation{Department of Physics and Astronomy, University of Victoria,
Victoria, BC V8P 5C2, Canada}
\affiliation{Perimeter Institute for Theoretical Physics, Waterloo, ON N2J 2W9,
Canada}

\author{Rana X Adhikari}
\affiliation{California Institute of Technology, Pasadena, CA 91125, USA}

\date{\today}

\begin{abstract}
\noindent
While global cosmological and local galactic abundance of dark matter is well established, its identity, physical size and composition remain a mystery.
In this paper, we analyze an important question of dark matter detectability through its gravitational interaction, using current and next generation gravitational-wave observatories to look for macroscopic (kilogram-scale or larger) objects.
Keeping the size of the dark matter objects to be smaller than the physical dimensions of the detectors, and keeping their mass as free parameters, we derive the expected event rates.
For favorable choice of mass, we find that dark matter interactions could be detected in space-based detectors such as LISA at a rate of one per ten years.
We then assume the existence of an additional Yukawa force between dark matter and regular matter.
By choosing the range of the force to be comparable to the size of the detectors, we derive the levels of sensitivity to such a new force, which exceeds the sensitivity of other probes in a wide range of parameters.
For sufficiently large Yukawa coupling strength, the rate of dark matter events can then exceed 10 per year for both ground- and space-based detectors.
Thus, gravitational-wave observatories can make an important contribution to a global effort of searching for non-gravitational interactions of dark matter.
\end{abstract}

\maketitle

\input{intro}

\input{DetectionStatistics}

\input{Stochastic}

\textit{Acknowledgements.}---The work of  MP is supported in part by NSERC, Canada, and research at the Perimeter Institute
is supported in part by the Government of Canada through NSERC and by the Province of Ontario through MEDT.
EDH, TC, VVF, and RXA are supported in part by the NSF under award PHY-0757058.

\bibliography{LigoDMbiblio,RanaGWrefs}

\end{document}

%% file: intro.tex
\textit{Introduction.}---There is overwhelming evidence that the Universe is dominated by dark energy (DE) and dark matter (DM),
which together comprise about $95\%$ of the cosmological critical energy  density
$\rho_\text{c}\times c^2 \simeq 5\,\text{keV/cm}^3$~\cite{Agashe:2014kda}.
Thus far, all the evidence comes from the gravitational influences of DE and DM on regular matter built from the Standard Model (SM) particles and fields.
The concentration of DM is enhanced around collapsed cosmic structures, such as galaxies and clusters of galaxies, where it exceeds its cosmological average by several orders of magnitude.
In particular, the energy density of dark matter in the Milky Way close to the location of the solar system has been determined to be about $0.39\,\text{GeV/cm}^3$~\cite{Catena:2011kv}.
The observed DM behavior is consistent with its being ``cold'', which implies a certain
Maxwellian-type velocity distribution, with an rms velocity of about $270\,\text{km/s}$ inside the Milky Way.
This random motion is superimposed on the ${\sim}\,220\,\text{km/s}$ constant velocity of the Sun relative to galactic center, so that there is a significant asymmetry in the flux of dark matter for an observer on earth.

Since all information on DM comes from its gravitational interactions, its composition and properties remain unknown. Among the most
important questions that do not have any direct observational answers are the following:
\begin{itemize}

\item What is the relation of DM to the visible matter of the SM? Is there any new interaction that
  supplements gravity and acts between DM and regular atoms?

\item Is DM elementary or composite?

\item What is the physical size of the DM objects and their mass?
\end{itemize}

In many particle physics models, DM is elementary and can be represented either by massive particles ({\em e.g.}, related to the lightest supersymmetric partners of SM particles), or by light fields ({\em e.g.} QCD axions).
Extensive research aimed at the direct detection of DM has advanced the sensitivity to elementary DM interacting with atoms, nuclei and electromagnetic fields.
It has produced  bounds on {\em e.g.} weak-scale DM interacting with nuclei~\cite{Cushman:2013zza}, but so far has not led to any  answers to the above questions.
While the next generation of such experimental efforts may bring positive results, it is important to {\em widen} the DM search program using the multi-probe approach with sensitive instruments.

In this Letter, we investigate the use of gravitational-wave observatories as detectors of dark matter via gravitational interaction of DM objects with the detectors' test masses.
The gravitational interaction is the only guaranteed interaction between DM and SM, and therefore it is important to investigate the prospects of a detection based only on gravitational interaction.
Moreover, 
we will study detection based on possible additional interactions -- modeled as a Yukawa potential -- between dark matter and the particles of the standard model.

\textit{The model of macroscopic DM.}---The discussion of macroscopic-size dark matter was traditionally oriented towards
the massive compact halo objets (MACHOs) and primordial black holes. The range of suggested masses
for these candidates starts from rather large values, $M>10^{14}\,\text{g}$ \cite{Yoo:2003fr,Capela:2013yf}. This mass range
influenced early discussions on a possible use of space-based gravitational-wave inteferometers in
search for dark matter~\cite{Seto:2004zu,Adams:2004pk}. For primoridal black holes, the range below $10^{14}\,\text{g}$ is disfavored due to
Hawking evaporation~\cite{Hawking:1974sw} shortening the lifetime below the age of the Universe.
Going away from the black hole candidates, one faces a much broader spectrum of macroscopically sized DM candidates~\cite{Pospelov:2012mt,Derevianko:2013oaa,Stadnik:2014tta, 2015arXiv151206165G}.
In particular, if sufficiently complex, dark sectors can possess stable topological monopoles~\cite{Hooft:1974qc,Polyakov:1974ek}, or non-topological defects, such as Q-balls~\cite{Coleman:1985ki}. 
Given the unknown properties of the dark sector, the mass range for such DM objects can be almost arbitrary, and their required cosmological abundance can be acheived via the so-called Kibble--Zurek mechanisms\,\cite{Zurek:1985qw}.
Microscopic particle-type DM can form objects much smaller than galactic size, also known as clumps. The size and mass density of such objects may widely differ depending on DM properties, and the cosmological history. 

For the purpose of this study, we will assume that DM consists of macroscopic objects of a certain transverse radius $r_\text{DM}$ and mass $M_\text{DM}$.
The mass $M_\text{DM}$ determines the average distance between the DM objects, and the frequency of encounters. 
Introducing the number density of galactic DM objects, $n_\text{DM} \equiv L^{-3}$, we obtain the following relation between the mass and the characteristic distance between the DM objects,
\begin{equation}
\rho_\text{DM} = M_\text{DM} n_\text{DM} ~~\Longrightarrow  ~~\frac{L}{10^4~\text{km} } \simeq 1.2\times\left(\frac{M_\text{DM}}{1 ~\text{kg}}\right)^{1/3},
\end{equation}
where $\rho_\text{DM}$ is DM mass density, $\rho_\text{DM}\times c^2 \simeq 0.39\,\text{GeV/cm}^3$.

This distance can be directly related to the effective flux of DM, and the frequency of close 
encounters. For a fiducial choice of $M_\text{DM}$ of 1\,kg, the effective flux of DM is 
$\Phi_\text{DM} \sim n_\text{DM} v_\text{DM} \sim 3\times 10^{-10}\,\text{km}^{-2}\,\text{s}^{-1}$,
and one can expect one DM object per year to pass the detector with an impact parameter of 10\,km.
This is commensurate with the actual physical size of the interferometer arms of existing graviational-wave detectors such as LIGO~\cite{2016arXiv160203838T}, and if the interaction between the DM objects and atoms, which the gravitating masses of LIGO are made of, is strong enough, such passage could in principle be detected.
The generalization to other types of defects (strings and/or domain walls) is also possible~\cite{Pospelov:2012mt,Jaeckel:2016jlh}.

What kind of interaction could one expect to have between the DM and SM? Besides 
purely gravitational interaction, the number of possibilities is quite large\,\cite{Derevianko:2013oaa}. In this Letter we will consider additional Yukawa interaction introduced by
the exchange of a light scalar, vector or tensor particle with mass $m_\phi \equiv \lambda^{-1}\times (\hbar/c)$.
Combining Yukawa and gravitational interactions, we write the non-relativistic potential between the two compact objects, separated at distance $r$ ($r > r_\text{DM}$), as follows:
\begin{eqnarray}
\label{ansatz}
V_{i-j} = - M_{i}M_j \frac{G_\text{N}}{r}\Big(1 +(-1)^s~\delta_i\delta_j \exp[-r/\lambda] \Big)
\\\nonumber \text{where }~ i,j =\text{SM,DM.} ~~~~~~~~~~~~
\end{eqnarray}
This equation assumes that the potential scales with the mass of the object ({\em e.g.} $\phi T_\mu^\mu$ coupling in the scalar case), and the corresponding couplings are parametrized in units of the standard gravitational coupling by the dimensionless numbers $\delta_\text{SM}$ and $\delta_\text{DM}$.
$(-1)^s$ is equal to $+1$ for scalar and tensor exchange, and $-1$ for vector exchange.
Moreover, we shall assume that the range of the force and the physical size of the detectors (LIGO) are much larger than the size of the DM objects, but smaller than the average distance between them,
\begin{equation}
r_\text{DM} \ll l_\text{LIGO}, \lambda \ll L,
\label{hierarchy}
\end{equation}
which significantly simplifies the analysis.

Extensive tests of the gravitational force, $V_\text{SM-SM}$, have set stringent constraints on $\delta_\text{SM}$
as a function of $\lambda$\,\cite{Schlamminger:2007ht}. Thus, for $\lambda \sim 1$\,km, $|\delta_\text{SM} |< 10^{-3}$.
At the same time, the coupling of this Yukawa force to DM can be many orders of magnitude stronger.
The main constraint on $\delta_\text{DM}$ comes from the influence of DM self-interaction on structure formation\,\cite{Spergel:1999mh}
and on the dynamics of cluster collisions\,\cite{Harvey:2015hha}. Since the range of the force is assumed to be less than $L$, only pair-wise collisions
are important. The momentum-exchange cross section can be easily calculated with the use of the inequalities in Eq.~(\ref{hierarchy}). To logarithmic accuracy it is given by
\be
\sigma_\text{DM-DM} = 16\,\pi \times \frac{G_\text{N}^2\,M_\text{DM}^2\,\delta_\text{DM}^4} {v_\text{DM}^4}\times \log\left[ \frac{\lambda}{r_\text{DM}} \right].
\ee
At $v_\text{DM} \sim 10^{-3} c$, there is a typical constraint on the cross section,
$\sigma_\text{DM-DM}/M_\text{DM} \lesssim 1 \text{cm}^2/\text{g}$,
which translates to the following limit on the value of the DM Yukawa coupling,
\be
\label{deltaDM}
|\delta_\text{DM}| \lesssim 5\times 10^{9} \times \left( \frac{1~\text{kg}}{M_\text{DM}} \right)^{1/4}.
\ee
In deriving this limit, we set the value of the logarithm to 5.

It is important to emphasize that {\em saturating} this bound 
may alleviate some problems of cold DM scenario that emerge when 
observations are compared to numerical simulations. 
Self-interaction helps to cure the problem of cold DM overly-dense central regions of
dwarf galaxies predicted in simulations\,\cite{BoylanKolchin:2011de}, as DM self-scattering reduces the DM densities in the central regions
relative to non-interacting case
(see {\em e.g.} \cite{Vogelsberger:2012ku}). Therefore, $  |\delta_\text{DM}|  \gg  1$
represents a phenomenologically motivated choice. Taking two limits on $\delta_i$ together, one can conclude that at $r<\lambda$ the strength
of DM-SM interaction, $  |\delta_\text{DM} \delta_\text{SM} |$, can exceed gravity by up to seven orders of magnitude. 
One microscopic realization of $|\delta_\text{DM}|\gg|  \delta_\text{SM} |$ possibility would be a new scalar force with 
reasonably strong coupling to DM, and reduced coupling to the SM mediated {\em e.g.}
via the Higgs portal\,\cite{Piazza:2010ye}.

%% file: DetectionStatistics.tex
\textit{Macroscopic DM detection.}---We perform several Monte Carlo simulations in order to characterize the rate of discrete DM interaction events with laser interferometers.
We first consider the case of a single Advanced LIGO detector~\cite{aLIGO:CQG} operating at full sensitivity.
A worldwide network of such kilometer-scale laser interferometers will come into operation during the next several years~\cite{aLIGO:CQG, Virgo:Review2012, Akutsu:2015-05-11T00:00:00:1742-6596:12016}.
Future terrestrial~\cite{ET:2014, Dwyer:2015df} and space-based detectors~\cite{Sesana:2014jk} have also been planned.
We therefore also consider the case of a single LISA-type detector.

\begin{figure}[tbp]
    \centering
    \includegraphics[width=\columnwidth]{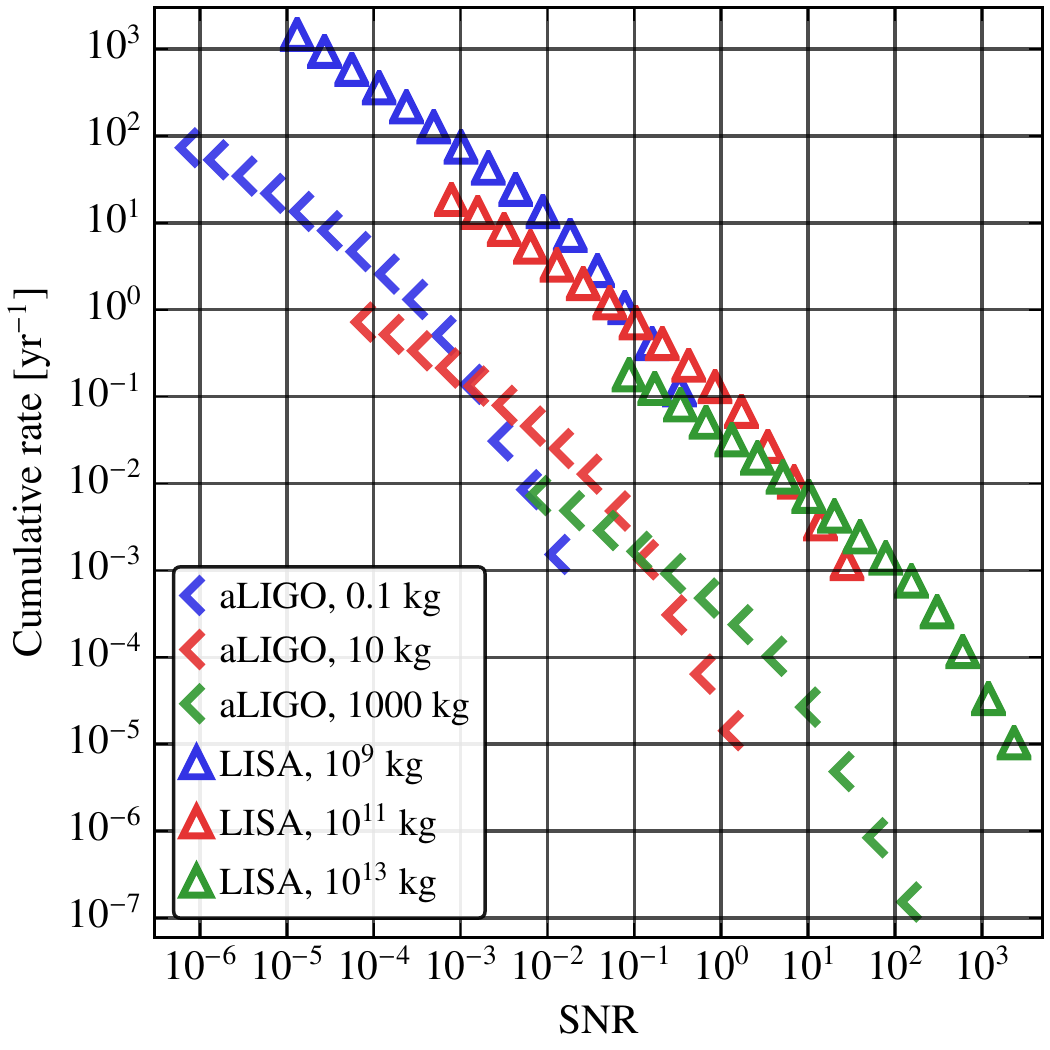}
    \caption{(color online).
    Cumulative event rate for gravitational interactions in a single Advanced LIGO detector and in a single LISA detector.
    }
    \label{fig:NewtonianRate}
\end{figure}

We model the distribution of DM in the galaxy as objects of mass $M$, with a uniform density in the solar system of $\rho_\text{DM} = (0.39\,\text{GeV}/c^2)/\text{cm}^3$, and a randomly directed velocity $\mathbf{v}$ whose magnitude is distributed according to a combination of the galaxy-frame DM velocity (270\,km/s rms) and the speed of the solar system through the galaxy (220\,km/s).
As the DM object (or undisrupted clump of DM) 
passes by the detector, it produces an acceleration $\mathbf{a}^{(k)}(t)$ of the detector's $k$th test mass (four in the case of LIGO, conventionally labeled as IX, IY, EX, and EY).
The acceleration is determined by the gradient of Eq.~(\ref{ansatz}) with $i = \text{SM}$ and $j = \text{DM}$.
The detector's GW channel reads out the differential acceleration
$a(t) = \bigl[a^{(\text{EX})}_x(t) - a^{(\text{IX})}_x(t)\bigr] -
\bigl[a^{(\text{EY})}_y(t) - a^{(\text{IY})}_y(t)\bigr]$~\cite{Saulson1994}.
We assume that the signal of this event can be optimally recovered from the detector's time stream using matched filtering; \emph{i.e.}, the signal-to-noise ratio (SNR) is $\varrho = \left[4\int_0^\infty \mathrm{d}\!f\, |a(f)|^2 / S_{nn}(f)\right]^{1/2}$, where $a(f)$ is the Fourier transform of $a(t)$ and $S_{nn}(f)$ is the power spectral density (PSD) of the detector's acceleration noise $n(t)$~\cite{Maggiore2007}.

In addition to simulating several DM masses for each detector, we also vary the coupling $g = \delta_\text{SM}\delta_\text{DM}$ and the screening $\lambda$, as defined in Eq.~(\ref{ansatz}).
The Newtonian case ($g=0$) has already been analyzed analytically in the context of primordial black hole detection with LISA~\cite{Seto:2004zu}, in the limits $b \ll \ell$ (the ``close-approach'' limit) and $b \gg \ell$ (the ``tidal'' limit), in both cases assuming a flat detector noise PSD and normal incidence of the masses to the detector plane.

\begin{figure*}[tbp]
    \centering
    \includegraphics[width=0.8\textwidth]{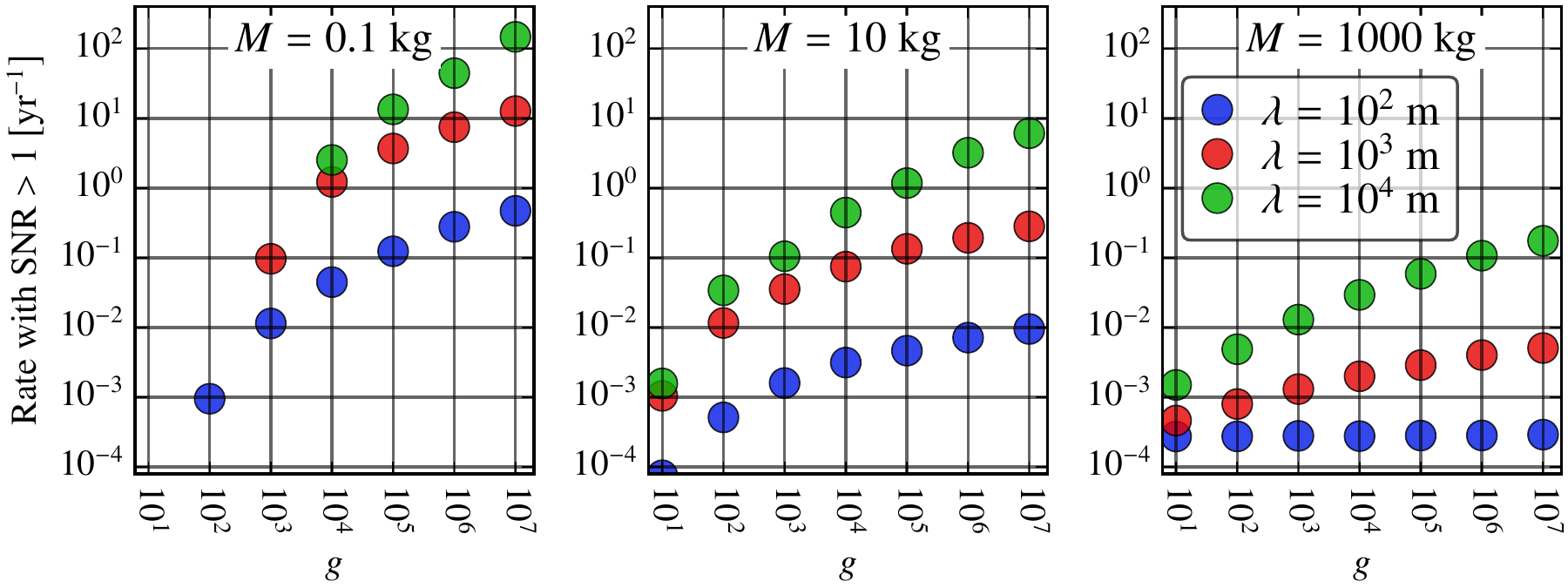}
    \caption{(color online).
    Event rate $\dot\eta(1)$ for non-SM interactions in a single Advanced LIGO detector, as a function of coupling $g = \delta_\text{SM}\delta_\text{DM}$ and screening length $\lambda$.}
    \label{fig:LIGOYukawaRate}
\end{figure*}

\begin{figure*}[tbp]
    \centering
    \includegraphics[width=0.8\textwidth]{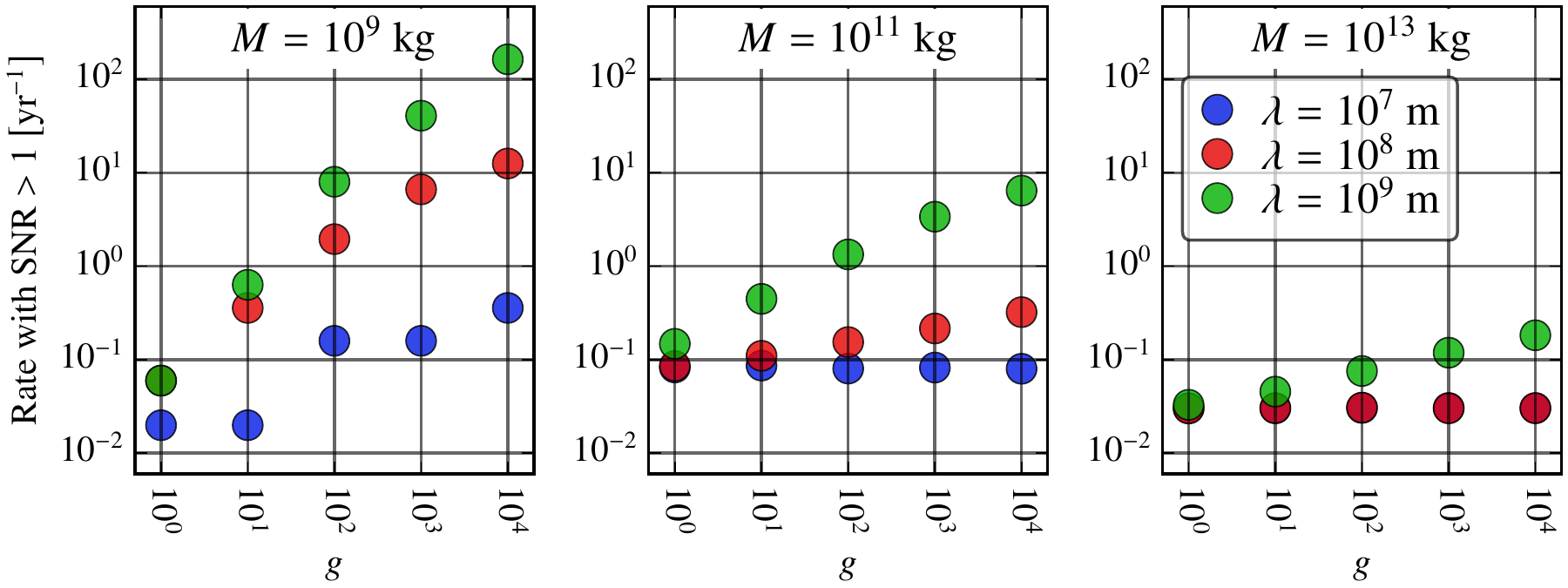}
    \caption{(color online).
    Event rate $\dot\eta(1)$ for non-SM interactions in a single LISA detector, as a function of coupling $g = \delta_\text{SM}\delta_\text{DM}$ and screening length $\lambda$.}
    \label{fig:LISAYukawaRate}
\end{figure*}

We then compute the cumulative rate function $\dot\eta(\varrho)$, which gives the number of events per year with SNR above $\varrho$.
In Fig.~\ref{fig:NewtonianRate} we plot the detector interaction rates assuming a Newtonian coupling.
In Figs.~\ref{fig:LIGOYukawaRate} and \ref{fig:LISAYukawaRate} we show how $\dot\eta$ is enhanced if the SM--DM interaction follows a Yukawa force law.
The ability of LIGO and LISA to place constraints on $g$ and $\lambda$ depends on the mass of DM object; in both cases, the smallest masses considered (0.1\,kg for LIGO, $10^9$\,kg for LISA) allow for the most sensitivity to $\{g,\lambda\}$ parameter space. 
If we choose $\delta_\text{SM}$ close to the existing bounds, 
and $\delta_\text{DM}$ to saturate (\ref{deltaDM}), then the rate of loud encounters can exceed $\mathcal{O}(10)$ per year. 

To confidently claim detection, a DM signal must be distinguished from glitches and other detector artifacts.
One strategy is to look for DM signals using two (nearly) co-located detectors.
The Advanced LIGO detectors as currently built are not co-located, though the Hanford facility did house two co-located Initial LIGO detectors.
Some of the plans for LISA-like space missions involve three co-located detectors~\cite{NGO:YellowBook}.

%% file: Stochastic.tex
\textit{Stochastic DM detection.}---In addition to single, loud DM events, we alternatively consider the case of a stochastic DM background due to a population of lighter, individually unresolvable DM objects.
Cross-correlating the outputs of GW detectors placed at remote points on the earth, reduces vastly the event rate.
In order to place best-case limits on our ability to detect such a signal, we consider only the case of two identical, colocated, and coaligned detectors whose noise is stationary, Gaussian, and independent.

Assuming the DM background $a(t)$ is independent of, and much weaker than, the detector noises $n_1(t)$ and $n_2(t)$, the optimal SNR is $\left[2T\int_0^\infty \mathrm{d}\!f\; S_{aa}(f)^2/S_{nn}(f)^2\right]^{1/2}$, where $S_{aa}(f)$ is the PSD of $a$, $T$ is the observing time, and we assume $S_{n_1 n_1} = S_{n_2 n_2} \equiv S_{nn}$.
We find that a Newtonian DM background is undetectable after $T = 5$\,years for the DM masses considered: for LIGO, masses of $10^{-9}$--$10^{-7}$\,kg result in optimal SNRs of 0.3--$5\times10^{-17}$; for LISA, masses of $10^6$, $10^7$, and $10^8$\,kg result in optimal SNRs of $9\times10^{-7}$, $4\times10^{-6}$, and $1.4\times10^{-4}$, respectively.
However, for $g \gg 1$, we have $S_{aa} \propto |g|^2$, and hence the SNR increases with $|g|^2$.
Therefore, LISA could detect a stochastic background from Yukawa interaction of DM clumps with mass $10^8$\,kg provided $|g| \gtrsim 10^2$, or clumps
with mass $10^6$\,kg provided $|g| \gtrsim 10^3$.